\documentclass[11pt,a4paper]{article}
\usepackage{amssymb}
\usepackage{jcappub}

\pdfoutput=1
\affiliation{Pushkov Institute of Terrestrial Magnetism, Ionosphere and Radiowave
Propagation of the Russian Academy of Science (IZMIRAN), \\
108840 Troitsk, Moscow, Russia}
\emailAdd{leinson@yandex.ru}
\abstract{To constrain the allowed range for the axion decay constant $f_{a}$ or, equivalently, for the axion mass $m_{a}$, we consider the cooling of a neutron star with strong proton superfluidity and normal (non-superfluid) neutrons inside its core and without strong magnetic field, by analogy with the observed supernova remnant in HESS J1731-347. For this specific case, we demonstrate that after the thermal relaxation is over, the hydrostatic structure of the neutron star can be well described with the aid of solution of Einstein field equations, applied to a sphere of fluid in hydrostatic equilibrium, derived by Tolman.  The internal temperature of the neutron star is calculated assuming that the cooling occurs dominantly due to production of neutrino pairs and axions in the nn-bremsstrahlung.
To impose a constraint to the axion decay constant the fact is used that the currently observed neutron star surface temperature does not deviate from the neutrino cooling scenario. For the KSVZ-axion model we find that $f_{a}>1.\,9\times 10^{8}$ GeV, while for the DFSZ-axion model we obtain $f_{a}>4.\,\allowbreak 7\times 10^{9}$ GeV.}

\begin{document}

\title{Constraints on axions from neutron star in HESS J1731-347}
\author{Lev B. Leinson}
\maketitle

\flushbottom

\section{Introduction}

\label{sec:intro}

Axions are hypothetical Nambu-Goldstone-bosons associated with the
spontaneously broken Peccei-Quinn symmetry that have been suggested as a
solution of the CP-violation problem in the strong interactions \cite%
{P77,W78,Wl78}. The scale of symmetry-breaking, which is also called the
axion decay constant $f_{a}$, is left undetermined in the theory. Though
axions arise as Nambu-Goldstone bosons and thus must be fundamentally
massless their interaction with gluons induces their mixing with neutral
pions. Axions thereby acquire a small mass which is inverse proportional to
the scale of symmetry-breaking. \cite{BT78,KSS78,PBY87,GKR86}: 
\begin{equation}
m_{a}=\frac{z^{1/2}}{1+z}\frac{f_{\pi }m_{\pi }}{f_{a}}\;,  \label{maz}
\end{equation}%
where $z=m_{u}/m_{d}\approx 0.56$ is canonically used \cite%
{Gasser198277,Leutwyler:1996qg}, although it could vary in the range $%
z=0.3-0.6$ \cite{y06}, the pion mass $m_{\pi }=135$ MeV, its decay constant
is $f_{\pi }=92$ MeV. Since axions are a plausible candidate for the cold
dark matter of the universe, a reasonable estimate of the axion mass $m_{a}$
(or, equivalently, the axion decay constant) represents much interest.

Two types of axion models are known: the Kim-Shifman-Weinstein-Zakharov
(hadronic) -- KSVZ model \cite{KSVZ,KSVZ1}, where the axion interacts only
with photons and hadrons, and the Dean-Fischler-Srednitsky-Zhitnitsky --
DFSZ model \cite{DFSZ,DFSZ1} involving the additional axion coupling to the
charged leptons. For a general review on axion physics see, e.g., \cite%
{Kim,Cheng}. The axion phenomenology, in particular in relation with the
astrophysical processes, is largely discussed in \cite%
{VZKC,R,R90,T90,R99,R08}.

The axion interaction with fermions $j$ has a derivative structure. We will
focus in the axion interaction with non-relativistic neutrons. The
corresponding Hamiltonian density can be written in the form (we use natural
units, $\hbar =c=k_{B}=1$): 
\begin{equation}
\mathcal{H}_{an}=\frac{\mathfrak{c}_{n}}{2f_{a}}\delta _{\mu i}\left( \Psi
^{+}\hat{\sigma}_{i}\Psi \right) \partial ^{\mu }a,  \label{Han}
\end{equation}%
where $\Psi $ is a neutron field with mass $m_{n}$, $\mathfrak{c}_{n}$ is a
model dependent numerical coefficient, and $\hat{\sigma}_{i}$ are the Pauli
spin matrices. The combination $g_{ann}=\mathfrak{c}_{n}m_{n}/f_{a}$ plays a
role of a Yukawa coupling. For nucleons, the dimensionless couplings $%
\mathfrak{c}_{n}$ are related by generalized Goldberger-Treiman relations to
nucleon axial-vector current matrix elements. A recent determination using
lattice QCD finds \cite{chvv16}: 
\begin{eqnarray*}
\mathfrak{c}_{n}^{\mathrm{KSVZ}} &=&-0.02(3), \\
\mathfrak{c}_{n}^{\mathrm{DFSZ}} &=&0.254-0.414\sin ^{2}{\beta }\pm 0.025,
\end{eqnarray*}%
where $\cot {\beta }$ is the ratio of the vacuum expectation values of the
two Higgs fields in the DFSZ model.

In order to constrain the permissible range for the axion mass $m_{a}$
various laboratory experiments, as well as astrophysical arguments, are in
use (see e.g. \cite{VZKC,M88,M89,omy96}). In order to avoid an "overclosed
universe" the Peccei-Quinn scale must obey an upper limit (a lower limit on
the axion mass). Currently \cite{AS83,DF83}, cosmological arguments give $%
m_{a}>10^{-5}$ eV. The most stringent upper limits on the axion mass derive
from astrophysics. Strength of the axion coupling with normal matter and
radiation is bounded by the condition that stellar-evolution lifetimes or
energy-loss rates not conflict with observation. Such arguments are normally
applied to the physics of supernova explosions, where the dominant energy
loss process is the emission of neutrino pairs and axions in the nucleon
bremsstrahlung. \cite{Br88,Bu88,R93,H00}. The requirement that stars must
not lose energy too effciently by emission of axions leads to a lower limit
on the Peccei-Quinn scale $f_{a}$ or, equivalently, to an upper limit on the
axion mass $m_{a}$. The limit from Supernova 1987A gives $m_{a}<0.01$ eV 
\cite{RS91,J96}. The transient behavior of the Cas A was studied in Refs. 
\cite{l14,hnyz18}, where axion emission was added to compensate for the
deficit of neutrino energy losses in order to reproduce the seeming\footnote{%
The data on Cas A rapid cooling is inconclusive, see \cite{ehshsypd13,rp18}}
rapid cooling of this object reported in \cite{H09,H10}. In works \cite%
{I84,U97,s16,s19} the thermal evolution of a cooling neutron star was
studied by including the axion emission in addition to neutrino energy
losses. The authors suggest the upper limits on the axion mass of order $%
m_{a}<0.06-0.3$ eV by comparing the theoretical curves with the ROSAT
observational data for three pulsars: PSR 1055-52, Geminga and PSR 0656+14.
Recently, even more sophisticated Markov-Chain Monte Carlo method has been
used \cite{brpr18} in numerical simulations of thermal evolution of the hot
young neutron star in the supernova remnant HESS J1731-347.

\section{Why XMMU J1732}

\label{sec:J1731}

From observations of the X-ray spectra of neutron stars, one can obtain an
estimate of the axion parameters if one makes a preliminary conclusion that
these X-rays are actually thermal and that heating mechanisms do not work
inside neutron stars. Therefore, to derive the astrophysical restrictions to
the axion coupling with nucleons it would be most convenient to choose a
neutron star observed as a central compact object that does not possess
magnetic activity and can be interpreted as a thermal emitting point-like
X-ray source which does not emit in other electromagnetic wavebands.

The most suitable to this can be considered the neutron star XMMU
J173203.3--344518 (hereafter XMMU J1732) which was discovered in x-ray
observations with XMM-Newton, Chandra, Suzaku, and Swift satellites \cite%
{hess09,tllyyyl10} as a point source near the center of HESS J1731-347
supernova remnant. This central compact object does not possess magnetic
activity, such as nonthermal (magnetospheric) radiation, and, apparently,
represents a young thermally radiating cooling neutron star. Lack of
magnetospheric and accretion phenomena potentially provides a view of the
surface of the star and its (effective) temperature. XMMU J1732 is
classified as a young, age $\simeq 27$ kyr, neutron star with a high surface
temperature, $T_{eff}\simeq 2.2\times 10^{6}$ K that makes it the hottest
known cooling neutron star, with likely a very weak magnetic field.

Thermal emission of XMMU J1732, can be interpreted as radiation from the
entire star surface with realistic mass and radius \cite{kps13}. The cooling
theory of neutron stars enables one to explain the exceptionally high
observable surface temperature of this star by assuming the presence of
strong proton superfluidity in the stellar core with normal (non-superfluid)
neutrons and the existence of the surface heat blanketing envelope which
almost fully consists of carbon \cite{kspysw15,okksy15}. As is well known,
the strong proton superfluidity suppresses all the neutrino emitting
reactions with proton participation (see e.g. \cite{y01,lp06}), therefore
the neutrino cooling of the neutron star takes place dominantly due to
neutrino pairs production in the nn-bremsstrahlung reaction.

In this connection, it is appropriate to recall that in a collision of
identical particles, dipole radiation in the vector channel of weak
interactions is absent and the emission of neutrino pairs occurs only due to
spin fluctuations owing to the collisions of neutrons in the medium. The
part of weak interaction responsible for this reaction in the long-wave limit%
\footnote{%
Since the neutron matter is strongly degenerate, the neutrino momentum $%
k\sim T$ is small compared to the neutron Fermi momentum.} is of the form%
\begin{equation}
\mathcal{H}_{\nu nn}=-\frac{G_{F}g_{A}}{2\sqrt{2}}\delta _{\mu i}\left( \Psi
^{+}\hat{\sigma}_{i}\Psi \right) l^{\mu },  \label{Hnu}
\end{equation}%
where $l^{\mu }=\bar{\nu}\gamma ^{\mu }\left( 1-\gamma _{5}\right) \nu $ is
the neutrino current, $G_{F}=1.166\times 10^{-5}$ GeV$^{-2}$ is the Fermi
coupling constant, $g_{A}\simeq 1.26$ is the neutral-current axial-vector
coupling constant of neutrons.

Since both the emission of neutrino pairs and axions are caused by
fluctuations of the spin density in the medium, the corresponding energy
losses due to the reaction of nn-bremsstrahlung differ from each other only
by the coupling constants and the phase volume of the outgoing particles.%
\footnote{%
The emitted neutrinos and axions are assumed freely escaping from the star.}
One can get the emissivity of neutrino pairs and axions in the form \cite%
{fm79,y01,hpr01,brpr18} (in ergs cm$^{-3}$s$^{-1}$):%
\begin{equation}
Q_{\nu }^{nn}=2.25\times 10^{20}\alpha _{nn}\!\beta _{nn}\left( \frac{%
m_{n}^{\ast }}{m_{n}}\right) ^{4}\!\!\left( \frac{n_{n}}{n_{0}}\right)
^{1/3}T_{9}^{8},\   \label{enu}
\end{equation}%
and 
\begin{equation}
Q_{a}^{nn}=3.\,\allowbreak 1\times 10^{38}g_{ann}^{2}\alpha _{nn}\!\beta
_{nn}\left( \frac{m_{n}^{\ast }}{m_{n}}\right) ^{4}\!\!\left( \frac{n_{n}}{%
n_{0}}\right) ^{1/3}\!T_{9}^{6}~,\ \   \label{eaex}
\end{equation}%
where $n_{n}$ is the number density of neutrons, $n_{0}=0.16\,fm^{-3}$ is
the nuclear saturation density, $m_{n}^{\ast }$ is the effective mass of a
neutron. The correction factors $\alpha _{nn}=0.59$ and $\beta _{nn}=0.56$
are introduced to account for numerous effects omitted in the analysis
(correlations, repulsive part of the nucleon nucleon interaction, etc.).

To constrain the allowed range for the axion decay constant $f_{a}$ or,
equivalently, for the axion mass $m_{a}$, we consider the cooling of a XMMU
J1732 type neutron star with strong proton superfluidity and normal
(non-superfluid) neutrons inside its core and without strong magnetic field.
The hydrostatic stellar structure is basically calculated by numerical
solving the Tolman-Oppenheimer-Volkoff (TOV) equations utilizing some
relevant equation of state (EOS) \cite{st83}. The cooling curves are,
however, known to be fairly independent of EOS, $M$, and $R$ for typical
theoretical neutron stars ($M\lesssim 1.8M_{\odot },R\sim 10-13~\mathrm{km}$%
), where the direct Urca process does not operate. The weak sensitivity of
the redshifted surface temperature $T_{s}^{\infty }(t)$ to variations of $M$
and $R$ for standard candles was first noted in \cite{pa92} and later
discussed in the literature (e.g. \cite{yp04} and references therein). \
Furthermore, as shown in \cite{lp01}, the profiles of mass-energy density ($%
\rho $), relative to central values ($\rho _{c}$), in neutron stars for
several EOSs demonstrate the behavior similar to a simple quadratic function.

In light of these arguments, we avail of the analytical solution discovered
by Tolman \cite{t39} who studied Einstein's field equations, applied to a
ball of fluid in hydrostatic equilibrium. For the case when the mass-energy
density$\rho $ changes quadratically, i.e.%
\begin{equation}
\rho =\rho _{c}[1-(r/R)^{2}],  \label{roroc}
\end{equation}%
the analytic solution to the general relativistic equations was found which
establishes simple analytical relations for the physical parameters of our
interest that are insensitive to the EOS. In terms of the variable $%
x=r^{2}/R^{2}$ and the compactness parameter $\beta =GM/R$, the assumption $%
\rho =\rho _{c}(1-x)$ results in $\rho _{c}=15\beta /(8\pi GR^{2})$. Making
use of space-like coordinates $r$, $\theta $ and $\phi $, and a time-like
coordinate $t$, such that the line element is described by the simple form 
\begin{equation}
ds^{2}=e^{2\Phi }dt^{2}-e^{\lambda }dr^{2}-r^{2}d\theta ^{2}-r^{2}\sin
^{2}\theta \,d\phi ^{2},  \label{ds}
\end{equation}
the solution of Einstein's equations for this density distribution can be
written as \cite{lp01}:%
\begin{equation}
e^{-\lambda }=1-\beta x(5-3x)\,,\qquad e^{2\Phi }=(1-5\beta /3)\cos ^{2}\phi
\,,  \label{grav}
\end{equation}%
\begin{equation}
P={\frac{c^{4}}{4\pi R^{2}G}}\left[ \sqrt{3\beta e^{-\lambda }}\tan \phi -{%
\frac{\beta }{2}}(5-3x)\right] \,,  \label{press}
\end{equation}%
\begin{equation}
n={\frac{(\rho c^{2}+P)}{m_{b}c^{2}}}{\frac{\cos \phi }{\cos \phi _{1}}}\,,\
\ \phi =(w_{1}-w)/2+\phi _{1}\,,  \label{nfi}
\end{equation}%
\begin{equation}
w=\log \left[ x-5/6+\sqrt{e^{-\lambda }/(3\beta )}\right] \,,\ \ \phi
_{c}=\phi (x=0)\,,  \label{wfi}
\end{equation}%
\begin{equation}
\phi _{1}=\phi (x=1)=\arctan \sqrt{\beta /[3(1-2\beta )]}\,,~\ w_{1}=w(x=1),
\label{fiw}
\end{equation}%
where $\Phi (r)$ is the gravitation potential, $P\left( r\right) $ is the
local pressure, and $n\left( r\right) $ is the number density of baryons.
This solution is scale-free, with the parameters $\beta $ and $\rho _{c}$
(or $M$ and $R$).

\begin{figure}[tbp]
\begin{center}
\includegraphics[width =1\textwidth]{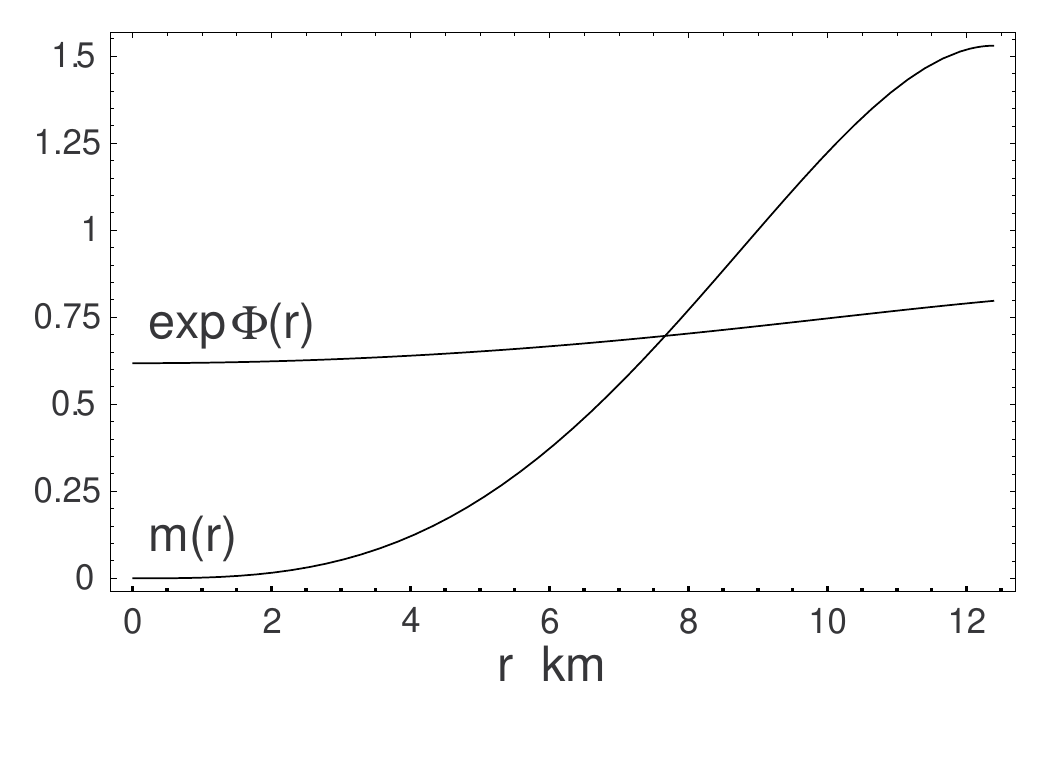}
\end{center}
\caption{The gravitational mass $m(r)\equiv M\left( r\right) /M_{\odot }$,
enclosed in a radius $r$, and the redshift factor $\exp \Phi (r)$ for a
netron star with the gravitational mass $M=1.53M_{\odot }$ and
circumferential radius $R=12.04$ km according to Eqs. (\protect\ref{grav})-(%
\protect\ref{fiw})}
\label{fig:fig1}
\end{figure}

For our estimates consider a neutron star with the gravitational mass $%
M=1.53M_{\odot }$ and circumferential radius $R=12.04$ km as is supposed to
the XMMU J1732 neutron star from observations \cite{kspysw15,okksy15}. For
this case the gravitational mass $m(r)\equiv M\left( r\right) /M_{\odot }$,
enclosed in a radius $r$, and the redshift factor $\exp \Phi (r)$, defined
in Eqs. (\ref{roroc}) and (\ref{grav}), are shown in Fig. \ref{fig:fig1}.

During the neutrino cooling stage the internal temperature of the neutron
star is controlled by the physical processes in its core. A thermally
relaxed star has an isothermal interior which extends from the center to the
heat-blanketing envelope located in the outer region at the matter density $%
\rho <10^{10}~$g cm$^{-3}$. Taking into account the effects of General
Relativity (e.g., \cite{thorne77}), isothermality means spatially constant
redshifted internal temperature 
\begin{equation}
\widetilde{T}=T(r)\exp (\Phi (r)),  \label{Tt}
\end{equation}%
while the local internal temperature $T(r)$, depends on the radial
coordinate $r$. At the neutrino cooling stage the cooling rate can be
described by the equation%
\begin{equation}
\frac{d\tilde{T}}{dt}=-l( \tilde{T}) ;~\ \ \ \ l( \tilde{T}) \equiv \frac{%
L^{\infty }(\widetilde{T})}{C( \tilde{T}) },  \label{ceq}
\end{equation}%
where $l( \tilde{T}) $ is the so-called cooling function -- the ratio of the
thermal energy losses per unit time $L^{\infty }(\widetilde{T})$ to the heat
capacity $C( \tilde{T}) $. This function is mainly determined by the
efficiency of the neutrino and axion production processes operating in the
neutron star core. We are interested in the case when the star possesses
very strong proton superfluidity which completely suppresses the modified
Urca process, np, and pp bremsstrahlung so that the star cools via the
neutrino and axion emission in nn collisions, i.e. 
\begin{equation}
l(\widetilde{T})=l_{\nu }(\widetilde{T})+l_{a}(\widetilde{T})  \label{ltot}
\end{equation}

At any time, the total luminosity of neutrino pairs $L_{\nu }^{\infty }(%
\widetilde{T})$ and axions $L_{a}^{\infty }(\widetilde{T})$ which can be
seen by distant observer depends on the redshifted internal temperature $%
\widetilde{T}$ and is given by the integrals over the neutron star core
volume (see e.g. \cite{yp04}): 
\begin{equation}
L_{\nu \left( a\right) }^{\infty }(\widetilde{T})=\int \mathrm{d}V\,Q_{\nu
\left( a\right) }^{nn}\left( T\left( r\right) \right) \exp (2\Phi (r)).
\label{Lnu}
\end{equation}%
In this expression, $T\left( r\right) $ is the local internal temperature, $%
r $ is radial coordinate, and $\Phi (r)$ is the gravitational potential that
determines gravitational redshift. We designate $\mathrm{d}V$ the element of
proper volume determined by the appropriate metric function:%
\begin{equation}
\mathrm{d}V=4\pi r^{2}\frac{dr}{\sqrt{1-r_{g}/r}},  \label{dv}
\end{equation}%
where $r_{g}=2GM\left( r\right) \approx 2.95M\left( r\right) /M_{\odot }$ km
is the Schwarzschild radius (here $G$ stands for gravitation constant, and $%
M\left( r\right) $ is the gravitational mass enclosed within radius $r$).

Inverting Eq.(\ref{Tt}) we obtain at any moment of time%
\begin{equation}
T(r)=\widetilde{T}\exp (-\Phi (r)).  \label{Tr}
\end{equation}%
Inserting this expression into Eq. (\ref{enu}) from Eq. (\ref{Lnu}) we get
(in $\mathrm{erg~s}^{-1}$)%
\begin{equation}
L_{\nu }^{\infty }(\widetilde{T})=7.\,\allowbreak 434\times 10^{19}\!\!\!\;\;%
\tilde{T}_{9}^{8}\int \mathrm{d}V\,\left( \frac{m_{n}^{\ast }}{m_{n}}\right)
^{4}\left( \frac{n_{n}}{n_{0}}\right) ^{1/3}\exp (-6\Phi (r)).  \label{LTnu}
\end{equation}%
Strong proton superfluidity fully suppresses also the proton heat capacity,
in this case we get%
\begin{equation}
C(\widetilde{T})=\int \mathrm{d}V\,c_{n}\left( T,\rho \right) ,  \label{C}
\end{equation}%
where the specific heat capacity of neutrons is $c_{n}=m_{n}^{\ast
}p_{Fn}k_{B}^{2}T/3$. One can can recast this expression to the form%
\begin{equation}
c_{n}=1.\,\allowbreak 61\times 10^{20}\frac{m_{n}^{\ast }}{m_{n}}\left( 
\frac{n_{n}}{n_{0}}\right) ^{1/3}T_{9}\ \ \ \ \ \mathrm{erg~}\mathrm{K}^{-1}%
\mathrm{cm}^{-3}.  \label{c}
\end{equation}%
Using Eq. (\ref{Tr}) we find%
\begin{equation}
C(\widetilde{T})=1.\,\allowbreak 61\times 10^{20}\;\tilde{T}_{9}\int \mathrm{%
d}V\,\frac{m_{n}^{\ast }}{m_{n}}\left( \frac{n_{n}}{n_{0}}\right) ^{1/3}\exp
(-\Phi (r)).  \label{CT}
\end{equation}%
Substituting Eqs. (\ref{LTnu}) and (\ref{CT}) into Eq. (\ref{ceq}) we obtain
the neutrino cooling function%
\begin{equation}
l_{\nu }\left( \tilde{T}\right) =q_{\nu }\tilde{T}^{7}  \label{l7}
\end{equation}%
with%
\begin{equation}
q_{\nu }=0.462\;\frac{\int \mathrm{d}V\,\left( m_{n}^{\ast }/m_{n}\right)
^{4}\left( n_{n}/n_{0}\right) ^{1/3}\exp (-6\Phi (r))}{\int \mathrm{d}%
V\,\left( m_{n}^{\ast }/m_{n}\right) \left( n_{n}/n_{0}\right) ^{1/3}\exp
(-\Phi (r))}  \label{q}
\end{equation}%
The axion cooling function is the ratio of the axion luminosity $%
L_{a}^{\infty }(\widetilde{T})$ to the heat capacity (\ref{C}). Using Eq. (%
\ref{eaex}), it can be found in the form 
\begin{equation}
l_{a}\left( \tilde{T}\right) =q_{a}\tilde{T}^{5},  \label{la}
\end{equation}%
where%
\begin{equation}
q_{a}=6.\,\allowbreak 38\times 10^{17}\!\!\!\;\;g_{ann}^{2}\,\frac{\int 
\mathrm{d}V\,\left( m_{n}^{\ast }/m_{n}\right) ^{4}\left( n_{n}/n_{0}\right)
^{1/3}\exp (-4\Phi (r))}{\int \mathrm{d}V\,\left( m_{n}^{\ast }/m_{n}\right)
\left( n_{n}/n_{0}\right) ^{1/3}\exp (-\Phi (r))}.  \label{qa}
\end{equation}

To evaluate $q_{\nu }$ and $q_{a}$, as defined in Eqs. (\ref{q}) and (\ref%
{qa}) one requires the distribution of neutron number density and the
effective mass over the radius of the neutron star's core in beta
equilibrium. To calculate these functions we employ the Walecka-type
relativistic model of baryon matter \cite{Serot}, where the baryons interact
via exchange of $\sigma $, $\omega $, and $\rho $ mesons (see details in 
\cite{l02,l02plb}). The functions $n_{n}\left( r\right) /n_{0}$ and $%
m_{n}^{\ast }\left( r\right) /m_{n}$ found in this way are shown in Fig. \ref%
{fig:fig2}.

\begin{figure}[tbp]
\begin{center}
\includegraphics[width =1\textwidth]{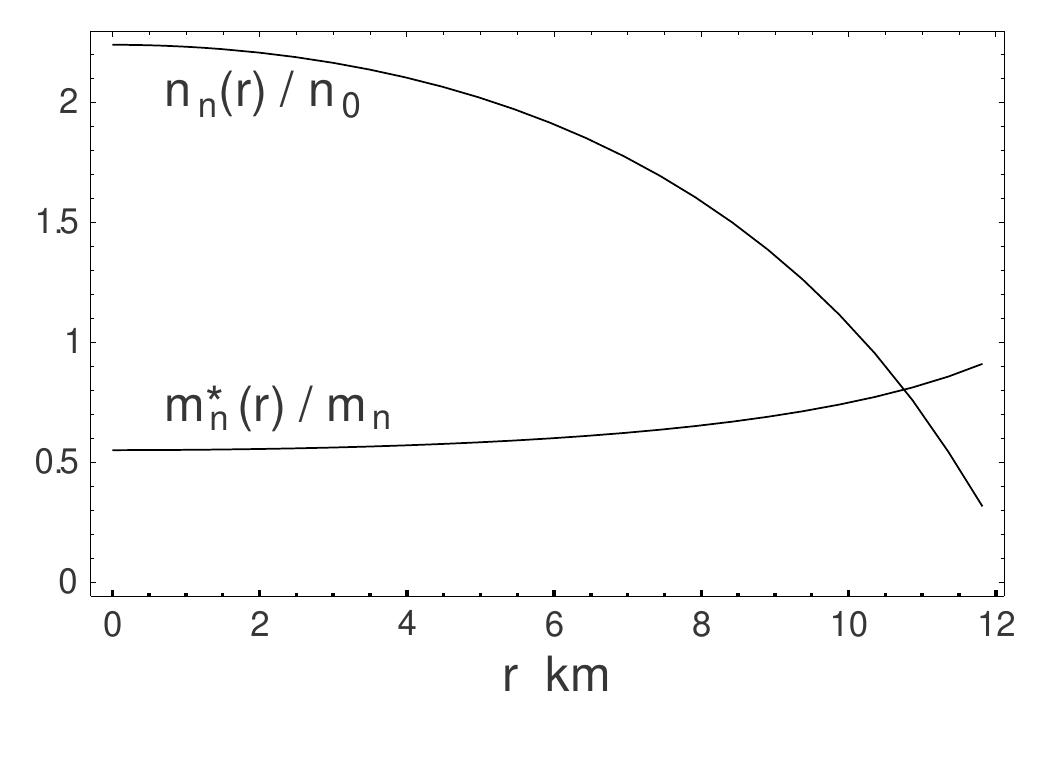}
\end{center}
\caption{Number density of neutrons $n_{n}\left( r\right) $ in units of $%
n_{0}=0.16\,$fm$^{-3}$ and the neutron effective mass $m^{\ast }/m_{n}$
against a distance from the center, as calculated according to $\protect%
\sigma ,\protect\omega ,\protect\rho $-model for a neutron star of total
mass $M=1.53M_{\odot }$ and radius $R=12.4\,$km.}
\label{fig:fig2}
\end{figure}

Trivial numerical integration over the core volume employing these function
gives 
\begin{equation}
q_{\nu }=0.823~\ \ \mathrm{K}^{-6}\mathrm{s}^{-1}  \label{qnu}
\end{equation}%
\ and 
\begin{equation}
q_{a}=4.483\times 10^{17}g_{nn}^{2}~\ \ \ \mathrm{K}^{-4}\mathrm{s}^{-1}.
\label{qA}
\end{equation}%
~

With making use of Eqs. (\ref{ltot}), (\ref{l7}), and (\ref{la}) the cooling
equation (\ref{ceq}) can be recast to%
\begin{equation}
\frac{d\tilde{T}}{dt}=-q_{\nu }\tilde{T}^{7}-q_{a}\tilde{T}^{5}.  \label{cra}
\end{equation}%
The temperature dependence of the cooling rates entering to this equation is
of a power law $l=q\tilde{T}^{n}$ with $n=7$ for slow neutrino cooling and
with $n=5$ for axions which corresponds to fast cooling process. This means
that the relative contribution of axions to the total energy losses, which
is minor at high temperatures, increases with the temperature lowering. As a
result the cooling curve must deviate from the neutrino cooling scenario
below some temperature. Obviously, the corresponding internal temperature $%
\widetilde{T}_{a}$ can be determined from the condition $l_{a}(\widetilde{T}%
)\sim l_{\nu }(\widetilde{T})$, giving%
\begin{equation}
q_{a}\sim q_{\nu }\tilde{T}_{a}^{2}.  \label{qanu} 
\end{equation}

Since, according to observations, the thermal radiation of XMMU J1732 can be
interpreted as neutrino radiation from nn-bremsstrahlung, it should be
assumed that axion losses are still relatively small in the current era,
i.e. $\tilde{T}^{2}\gg \tilde{T}_{a}^{2}$. Thus, the current observations in
a complex with the adopted cooling scenario allow us to conclude that%
\begin{equation}
q_{a}\ll q_{\nu }\tilde{T}_{9}^{2},  \label{T2}
\end{equation}%
where $\ \tilde{T}_{9}=\tilde{T}/10^{9}\mathrm{K}$ is the current internal
temperature of the neutron star, as it is seen by a distant observer.

\section{Observed temperature}

\label{sec:observation}

The internal temperature, indicated in Eq. (\ref{T2}), is virtually 
inaccessible for direct observations. Indeed, a thermally-relaxed star 
has an isothermal interior that extends from the center to a thin 
heat-blanketing envelope, where the main temperature gradient occurs.
Following \cite{gs80} one can adopt that the isothermal region is
restricted by the condition $\rho >\rho \left( r_{\mathrm{b}}\right)
=10^{10} $ \textrm{g cm}$^{-3}$. According to Eq. (\ref{Tr}) at the bottom
of heat-blanketing envelope, $r=r_{\mathrm{b}}$, one has $T_{\mathrm{b}%
}\equiv T(r_{\mathrm{b}})=\widetilde{T}\exp \left( -\Phi (r_{\mathrm{b}%
})\right) $. Since the heat-blanketing envelope is very thin one can put $%
\Phi (r_{\mathrm{b}})\simeq \Phi (R)$, thus obtaining 
\begin{equation}
\widetilde{T}=T_{\mathrm{b}}\sqrt{1-r_{g}/R}  \label{Tb}
\end{equation}

To relate the internal temperature of the neutron star to its surface
temperature $T_{\mathrm{eff}}$, which is available for observations we employ 
the fit for $T_{\mathrm{eff}}$-$T_{\mathrm{b}}$ relation from 
Ref. \cite{bpy16} for a carbon-iron heat blanketing envelope.
In this fit, the amount of carbon in the envelope, $\Delta M_C$, is parametrized by
\begin{equation}
\eta \equiv g_{14}^{2}\Delta M_C/M, \label{eta}
\end{equation}%
where $M$ is the NS mass, and $g_{14}$ is the surface gravitational acceleration, 
in units of $10^{14}$ cm s$^{-2}$. 
The effective surface temperature of XMMU J1732,
as measured by a distant observer, is $T_{\mathrm{s}}^{\infty }=1.78$ MK
(see \cite{kspysw15,okksy15}), which corresponds to the temperature at the
surface $T_{\mathrm{eff}}=2.23$ MK.

\begin{figure}[tbp]
\begin{center}
\includegraphics[width =1\textwidth]{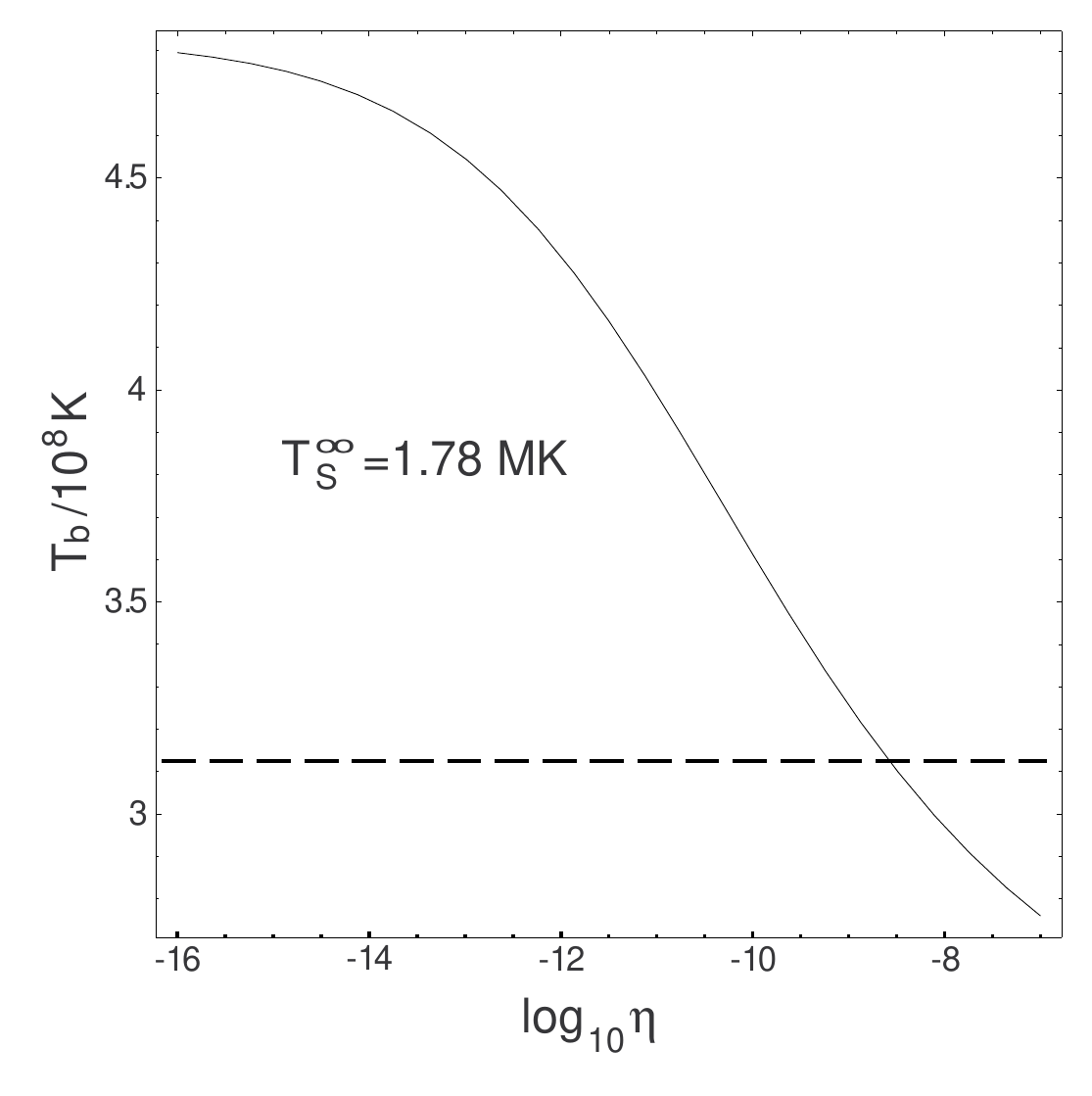}
\end{center}
\caption{The temperature $T_{\mathrm{b}}$ at the bottom of heat-blanketing 
envelope of a neutron star of mass $M=1.53M_{\odot }$ and
radius $12.4$ km for fixed temperature 
$T_{\mathrm{s}}^{\infty }=1.78$ MK, as a function of the parameter 
$\protect\eta $ which is proportional to the relative mass of carbon accreted 
to the neutron star surface.
The horizontal dashed line indicates the internal temperature, as calculated 
from Eqs. (\ref{Tr}) and (\ref{T9}).}
\label{fig:fig3}
\end{figure}

In Fig. \ref{fig:fig3}, we demonstrate the temperature $T_{\mathrm{b}}$ at
the bottom of heat-blanketing envelope as a function of the relative mass of
carbon accreted to the neutron star surface for fixed temperature $%
T_{\mathrm{s}}^{\infty }=1.78$ MK. At the fixed surface temperature the
internal temperature, reconstructed in this way, depends on the parameter $%
\eta $ and varies within $0.48\times 10^{9}\mathrm{K}>T_{\mathrm{b}%
}>0.275\times 10^{9}\mathrm{K}$ for the parameter $\eta $ ranging from $%
10^{-16}$ to $10^{-7}$. This domain of the temperature $T_{\mathrm{b}}$
corresponds to the redshifted internal temperature measured by a distant
observer within $0.22<\tilde{T}_{9}<0.38$. Substituting the largest of these
values into Eq. (\ref{T2}) and using Eqs. (\ref{qnu}) and (\ref{qA}), one
can safely write%
\begin{equation}
g_{ann}^{2}\ll 2.\,\allowbreak 6\times 10^{-19}  \label{top}
\end{equation}

This estimate can be improved if one takes into account that the upper limit
of the internal temperature $\tilde{T}_{9}\sim 0.38$ corresponds to a very
small mass of carbon in the envelope $\eta =10^{-16}$, while the
XMMU J1732's soft x-ray spectrum is best characterized by a thick carbon
atmosphere model \cite{kspysw15,okksy15}. A more accurate estimate of the
internal temperature of a neutron star can be made, given that under
conditions where axion radiation can be neglected, the neutrino cooling
process is described by the equation 
\begin{equation}
\frac{d\tilde{T}}{dt}=-q_{\nu }\tilde{T}^{7},  \label{dTnu}
\end{equation}%
which has the well-known analytic solution (e.g. \cite{plps04,yhshp11}):%
\begin{equation}
\tilde{T}_{9}=\left( 6q_{\nu }t/T_{\ast }\right) ^{-1/6},  \label{T9}
\end{equation}%
where $t$ has to be expressed in seconds. The normalization factor $T_{\ast
}=10^{9}$ is introduced to ensure the dimension of $q_{\nu }$ we used above.
Assuming a neutron star age of $27\,$kyr\textrm{\ }$\simeq 8.516\times
10^{11}\mathrm{s}$, from equation (\ref{T9}), we obtain $\tilde{T}_{9}=0.249$%
. This temperature corresponds to $T_{\mathrm{b}}=0.312\times 10^{9}$K,
which is shown in Fig. \ref{fig:fig3} by a horizontal line, and $\mathrm{%
log\,}\eta =-8.6$.

With $\tilde{T}_{9}=0.249$ the inequality (\ref{T2}) takes the form 
\begin{equation}
g_{ann}^{2}\ll 1.\,\allowbreak 1\times 10^{-19}.  \label{ineq}
\end{equation}%
Of course, both the above estimates are the same in order of magnitude,
which allows us to conclude that 
\begin{equation}
g_{ann}\lesssim 1.0\times 10^{-10}.  \label{gnn}
\end{equation}%
Thus, for the KSVZ-axion model, where $\mathfrak{c}_{n}=0.02$, we find that $%
f_{a}$ $>1.\,9\times 10^{8}$ GeV, while for the DFSZ-axion model with $%
\mathfrak{c}_{n}=0.5$ we obtain that $f_{a}$ $>4.\,\allowbreak 7\times
10^{9} $ GeV.

The axion mass is related to $f_{a}$ via 
\begin{equation}
m_{a}=0.60\,\text{eV}\,\frac{10^{7}\text{GeV}}{f_{a}}  \label{ma}
\end{equation}%
which allows one to convert the decay constant $f_{a}$ to the axion mass $%
m_{a}$. This yields 
\begin{equation}
m_{a}^{\mathrm{KSVZ}}<3\times 10^{-2}\,\,\text{eV},  \label{mKSVZ}
\end{equation}%
and 
\begin{equation}
m_{a}^{\mathrm{DFSZ}}<1.\,\allowbreak 3\times 10^{-3}\,\text{eV.}
\label{mDFSZ}
\end{equation}

\section{Results and discussion}

\label{sec:result}

In the absence of fast neutrino reactions similar to the direct Urca
processes, the cooling rate of a neutron star depends on the equation of
state of baryonic matter, mainly because of the strong dependence of the
superfluid energy gaps of protons and neutrons on the matter density. This
is not the case for a neutron star like XMMU J1732, which can be considered
in the limit of normal neutrons and strong proton superfluidity in the core 
\cite{kspysw15,okksy15}. With this in mind, we used the analytical solution
of the Einstein field equations applied to the balls of fluid in hydrostatic
equilibrium, found by Tolman \cite{t39} for the case when the mass-energy
density changes quadratically (qualitatively, this is typical for many
equations of state \cite{lp01}). This approach allows one to analytically
describe both neutrino and axion luminosities caused by nn-bremsstrahlung in
the hot core of a neutron star. The analytically calculated internal
temperature $\tilde{T}$ of a neutron star of mass $M=1.53M_{\odot }$ and
radius $12.4$ km at the age of $27$ kyr is in excellent agreement with the
current internal temperature of XMMU J1732 obtained by simulations of its
cooling by complex numerical methods in \cite{okksy15} and is in a
reasonable agreement with the observed surface temperature of this neutron
star.

Since the strong proton superfluidity suppress all the reactions with proton
participation, the cooling of the neutron star takes place dominantly due to
neutrino pairs (and axions) production in the nn-bremsstrahlung. The
temperature dependence of the cooling rates entering to the cooling equation
is of a power law $l_{\nu }=q_{\nu }\tilde{T}^{7}$ for slow neutrino cooling
and $l_{a}=q_{a}\tilde{T}^{5}$ for axions which corresponds to fast cooling
process. This means that the relative contribution of axions to the total
energy losses increases with the temperature lowering. Since currently the
cooling curve does not deviate from the neutrino cooling scenario we use the
condition $q_{a}\ll q_{\nu }\tilde{T}^{2}$ to impose of the restriction to
the axion decay constant $f_{a}$, thus obtaining $g_{ann}^{2}\ll
1.\,\allowbreak 1\times 10^{-19}$. For the KSVZ-axion model, where $%
\mathfrak{c}_{n}=0.02$, we find that $f_{a}>1.\,9\times 10^{8}$ GeV, while
for the DFSZ-axion model with $\mathfrak{c}_{n}=0.5$ we obtain $%
f_{a}>4.\,\allowbreak 7\times 10^{9}$ GeV.

Our estimate is in agreement with the widely discussed constraint on $%
g_{ann}^{2}+g_{app}^{2}\lesssim 3.6\times 10^{-19}$ obtained earlier \cite%
{k92,R08} using the duration of the neutrino signal observed from supernova
(SN) 1987A as well as the recent estimate $g_{ann}^{2}<7.7\times 10^{-20}$
obtained in \cite{brpr18}. Our estimate is in agreement also with, $%
g_{ann}^{2}<1.6\times 10^{-19}$ obtained within the KSVZ model assuming
neutrons are not paired in \cite{U97}, where the cooling of three
middle-aged ($10^{5}-5\times 10^{5}$ yrs) pulsars, PSR 0656 14, PSR 1055-52,
and PSR 0633 1748 (\textquotedblleft Geminga\textquotedblright ) was
considered, which all are in the photon cooling era.


\begin{thebibliography}{99}
\bibitem{P77} R. D. Peccei and H. R. Quinn, \emph{CP Conservation In The
Presence Of Instantons}, 1977 \emph{Phys. Rev. Lett.} \textbf{38} 1440
[doi:10.1103/PhysRevLett.38.1440].

\bibitem{W78} S. Weinberg, \emph{A New Light Boson?}, 1978 \emph{Phys. Rev.
Lett.} \textbf{40} 223 [doi:10.1103/PhysRevLett.40.223].

\bibitem{Wl78} F. Wilczek, \emph{Problem Of Strong P And T Invariance In The
Presence Of Instantons}, 1978 \emph{Phys. Rev. Lett.} \textbf{40} 279
[doi:10.1103/PhysRevLett.40.279].

\bibitem{BT78} W. A. Bardeen and S. H. H. Tye, \emph{Current Algebra Applied
to Properties of the Light Higgs Boson}, 1978 \emph{Phys. Lett.} \textbf{B 74%
} 229 [doi:10.1016/0370-2693(78)90560-9].

\bibitem{KSS78} J. Kandaswamy, P. Salomonson and J. Schechter, \emph{Mass of
the Axion}, 1978 \emph{Phys. Rev.} \textbf{D 17} 3051
[doi:10.1103/PhysRevD.17.3051].

\bibitem{PBY87} W. A. Bardeen, R.D. Peccei and T.Yanagida, \emph{Constraints
On Variant Axion Models}, 1987 \emph{Nucl. Phys.} \textbf{B 279} 401
[doi:10.1016/0550-3213(87)90003-4].

\bibitem{GKR86} H. Georgi, D.B. Kaplan and L.Randall, \emph{Manifesting the
Invisible Axion at Low-energies}, 1986 \emph{Phys. Lett.} \textbf{B 169} 73
[doi:10.1016/0370-2693(86)90688-X].

\bibitem{Gasser198277} J. Gasser and H. Leutwyler, \emph{Quark masses}, 1982 
\emph{Phys. Rep.} \textbf{87} 77 [doi:10.1016/0370-1573(82)90035-7].

\bibitem{Leutwyler:1996qg} H. Leutwyler, \emph{The ratios of the light quark
masses}, 1996 \emph{Phys. Lett.} \textbf{B 378} 313 [hep-ph/9602366]
[doi:10.1016/0370-2693(96)00386-3].

\bibitem{y06} W. M. Yao, et al.: (Particle Data Group),\emph{\ Review of
particle physics}, 2006 \emph{J. Phys.} \textbf{G 33} 1
[doi:10.1088/0954-3899/33/1/001].

\bibitem{KSVZ} J. E. Kim, \emph{Weak Interaction Singlet and Strong CP
Invariance}, 1979 \emph{Phys. Rev. Lett.} \textbf{43} 103
[doi:10.1103/PhysRevLett.43.103].

\bibitem{KSVZ1} M. A. Shifman, A. I. Vainshtein and V. I. Zakharov, \emph{%
Can Confinement Ensure Natural CP Invariance Of Strong Interactions?}, 1980 
\emph{Nucl. Phys.} \textbf{B 166} 493 [doi:10.1016/0550-3213(80)90209-6].

\bibitem{DFSZ} A. R. Zhitnitskii, \emph{Possible suppression of axion-hadron
interactions}, 1980 \emph{Sov. J. Nucl. Phys.} \textbf{31} 260.

\bibitem{DFSZ1} M. Dine, W. Fischler and M. Srednicki, \emph{A Simple
Solution To The Strong CP Problem With A Harmless Axion}, 1981 \emph{Phys.
Lett.} \textbf{B 104} 199 [doi:10.1016/0370-2693(81)90590-6].

\bibitem{Kim} J. E. Kim, \emph{Light Pseudoscalars, Particle Physics And
Cosmology}, 1987 \emph{Phys. Rept.} \textbf{150} 1
[doi:10.1016/0370-1573(87)90017-2].

\bibitem{Cheng} H. Y. Cheng, \emph{The Strong CP Problem Revisited}, 1988 
\emph{Phys. Rept.} \textbf{158} 1 [doi:10.1016/0370-1573(88)90135-4].

\bibitem{VZKC} M. I. Visotsskii, Ya. B. Zel'dovich, M. Yu. Khlopov and V. M.
Chechetkin, \emph{Some astrophysical limitations on axion mass}, 1978 \emph{%
Pis'ma v ZhETF} \textbf{27} 533; [English translation: 1978 \emph{JETP Lett.}
\textbf{27} 502].

\bibitem{R} G. G. Raffelt, \emph{Stars as Laboratories for Fundamental
Physics.} 1996 \emph{The Astrophysics of Neutrinos, Axions, and Other Weakly
Interacting Particles.}, The University of Chicago Press, U.S.A.

\bibitem{R90} G. G. Raffelt, \emph{Astrophysical methods to constrain axions
and other novel particle phenomena}, 1990 \emph{Phys. Rept.} \textbf{198} 1
[doi:10.1016/0370-1573(90)90054-6].

\bibitem{T90} M. S. Turner, \emph{Windows On The Axion}, 1990 \emph{Phys.
Rept.} \textbf{197} 67 [doi:10.1016/0370-1573(90)90172-X].

\bibitem{R99} G. G. Raffelt, \emph{Particle physics from stars}, 1999 \emph{%
Ann.Rev.Nucl.Part.Sci.} \textbf{49} 163 [hep-ph/9903472]
[doi:10.1146/annurev.nucl.49.1.163].

\bibitem{R08} G. G. Raffelt, \emph{Astrophysical axion bounds}, 2008 \emph{%
Lct. Notes Phys.} \textbf{741} 51, Springer-Verlag Berlin Heidelberg
[hep-ph/0611350].

\bibitem{chvv16} G. Grilli di Cortona, E. Hardy, J. Pardo Vega, and G.
Villadoro, \emph{The QCD axion, precisely}, 2016 \emph{J. High Energy Phys.} 
\textbf{01} 034 [arXiv:1511.02867] [doi:10.1007/JHEP01(2016)034].

\bibitem{M88} R. Mayle, et al. \emph{Constraints on Axions from SN 1987a},
1988 \emph{Phys. Lett.} \textbf{B 203} 188
[doi:10.1016/0370-2693(88)91595-X].

\bibitem{M89} R. Mayle, et al. \emph{Updated Constraints on Axions from SN
1987a}, 1989 \emph{Phys. Lett.} \textbf{B 219} 515
[doi:10.1016/0370-2693(89)91104-0].

\bibitem{omy96} I. Ogawa, S. Matsuki, K. Yamamoto, \emph{Interactions of
cosmic axions with Rydberg atoms in resonant cavities via the Primakoff
process}, 1996 \emph{Phys. Rev.} \textbf{D 53} 1740
[doi:10.1103/PhysRevD.53.R1740].

\bibitem{AS83} L. F. Abbott and P. Sikivie, \emph{A Cosmological Bound on
the Invisible Axion}, 1983 \emph{Phys. Lett.} \textbf{B 120} 133
[doi:10.1016/0370-2693(83)90638-X].

\bibitem{DF83} M. Dine and W. Fischler, \emph{The Not So Harmless Axion},
1983 \emph{Phys. Lett} \textbf{B 120} 137 [doi:10.1016/0370-2693(83)90639-1].

\bibitem{Br88} R. P. Brinkmann and M. S. Turner, \emph{Numerical Rates for
Nucleon-Nucleon Axion Bremsstrahlung}, 1988 \emph{Phys. Rev.} \textbf{D 38}
2338 [doi:10.1103/PhysRevD.38.2338].

\bibitem{Bu88} A. Burrows, M. S. Turner and R. P. Brinkmann, \emph{Axions
and SN 1987a}, 1989 \emph{Phys. Rev.} \textbf{D 39} 1020
[doi:10.1103/PhysRevD.39.1020].

\bibitem{R93} G. Raffelt and D. Seckel, \emph{A selfconsistent approach to
neutral current processes in supernova cores}, 1995 \emph{Phys. Rev.} 
\textbf{D 52} 1780 [astro-ph/9312019] [doi:10.1103/PhysRevD.52.1780].

\bibitem{H00} C. Hanhart, D.R. Phillips and S. Reddy, \emph{Neutrino and
axion emissivities of neutron stars from nucleon-nucleon scattering data},
2001 \emph{Phys. Lett.} \textbf{B 499} 9 [astro-ph/0003445]
[doi:10.1016/S0370-2693(00)01382-4].

\bibitem{RS91} G. Raffelt and D. Seckel, \emph{Multiple scattering
suppression of the bremsstrahlung emission of neutrinos and axions in
supernovae}, 1991 \emph{Phys. Rev. Lett.} \textbf{67} 2605
[doi:10.1103/PhysRevLett.67.2605].

\bibitem{J96} H.-T. Janka, W. Keil, G. Raffelt and D. Seckel, \emph{Nucleon
spin fluctuations and the supernova emission of neutrinos and axions}, 1996 
\emph{Phys. Rev. Lett.} \textbf{76} 2621 [astro-ph/9507023]
[doi:10.1103/PhysRevLett.76.2621].

\bibitem{l14} L. B. Leinson, \emph{Axion mass limit from observations of the
neutron star in Cassiopeia A}, 2014 \emph{J. Cosmol. Astropart. Phys.} 
\textbf{08} 031 [arXiv:1405.6873] [doi:10.1088/1475-7516/2014/08/031].

\bibitem{hnyz18} K. Hamaguchi, N. Nagata, K. Yanagi, and J. Zheng, \emph{%
Limit on the axion decay constant from the cooling neutron star in
Cassiopeia A}, 2018 \emph{Phys. Rev.} \textbf{D 98} 103015
[arXiv:1806.07151] [doi:10.1103/PhysRevD.98.103015].

\bibitem{ehshsypd13} K. G. Elshamouty, C. O. Heinke, G. R. Sivakoff, W. C.
G. Ho, P. S. Shternin, D. G. Yakovlev, D. J. Patnaude, and L. David, \emph{%
Measuring the Cooling of the Neutron Star in Cassiopeia A with all Chandra
X-Ray Observatory Detectors}, 2013 \emph{Astrophys. J.} \textbf{777} 22E
[arXiv:1306.3387] [doi:10.1088/0004-637X/777/1/22].

\bibitem{rp18} B. Posselt and G. G. Pavlov, \emph{Upper Limits on the Rapid
Cooling of the Central Compact Object in Cas A}, 2018 \emph{Astrophys. J.} 
\textbf{864} 135 [arXiv:1808.00531] [doi:10.3847/1538-4357/aad7fc].

\bibitem{H09} W. C. G. Ho, C. O. Heinke, \emph{A Neutron Star with a Carbon
Atmosphere in the Cassiopeia A Supernova Remnant}, 2009 \emph{Nature} 
\textbf{462} 71 [arXiv:0911.0672] [doi:10.1038/nature08525].

\bibitem{H10} C. O. Heinke, W. C. G. Ho, \emph{Direct Observation of the
Cooling of the Cassiopeia A Neutron Star}, 2010 \emph{Astrophys. J} \textbf{%
719} L167 [arXiv:1007.4719] [doi:10.1088/2041-8205/719/2/L167].

\bibitem{I84} N. Iwamoto, \emph{Axion Emission from Neutron Stars}, 1984 
\emph{Phys. Rev. Lett.} \textbf{53} 1198 [doi:10.1103/PhysRevLett.53.1198].

\bibitem{U97} H. Umeda, N. Iwamoto, S. Tsuruta, L. Qin and K. Nomoto, \emph{%
Axion mass limits from cooling neutron stars}, 1998 in Proceedings of the
International Conference on Neutron Stars and Pulsars, edited by N.
Shibazaki et al., Universal Academy Press (Frontiers science series no. 24),
p. 213 [astro-ph/9806337].

\bibitem{s16} A. Sedrakian, \emph{Axion cooling of neutron stars}, 2016 
\emph{Phys. Rev.} \textbf{D 93} 065044 [arXiv:1512.07828]
[doi:10.1103/PhysRevD.93.065044].

\bibitem{s19} A. Sedrakian, \emph{Axion cooling of neutron stars. II. Beyond
hadronic axions}, 2019 \emph{Phys. Rev.} \textbf{D 99} 043011
[arXiv:1810.00190] [doi:10.1103/PhysRevD.99.043011].

\bibitem{brpr18} M. V. Beznogov, E. Rrapaj, D. Page, and S. Reddy, \emph{%
Constraints on Axion-like Particles and Nucleon Pairing in Dense Matter from
the Hot Neutron Star in HESS J1731-347}, 2018 \emph{Phys. Rev.} \textbf{C 98}
035802 [arXiv:1806.07991] [doi:10.1103/PhysRevC.98.035802].

\bibitem{hess09} H. E. S. S. Collaboration, F. Acero, G. P{\"{u}}hlhofer, D.
Klochkov, N. Komin, Y. Gallant, D. Horns, and A. Santangelo, \emph{X- and
gamma-ray studies of HESS J1731-347 coincident with a newly discovered SNR}
[arXiv:0907.0642].

\bibitem{tllyyyl10} W. W. Tian, Z. Li, D. A. Leahy, J. Yang, X. J. Yang, R.
Yamazaki, and D. Lu, \emph{X-Ray Emission from HESS J1731-347/SNR G353.6-0.7
and Central Compact Source XMMS J173203-344518}, 2010 \emph{Astrophys. J.} 
\textbf{712} 790 [arXiv:0907.1684] [doi:10.1088/0004-637X/712/2/790].

\bibitem{kps13} D. Klochkov, G. P{\"{u}}hlhofer, V. Suleimanov, et al., 
\emph{A non-pulsating neutron star in the supernova remnant HESS
J1731-347/G353.6-0.7 with a carbon atmosphere}, 2013 \emph{Astron. Astrophys.%
} \textbf{556} A41 [arXiv:1307.1230] [doi:10.1051/0004-6361/201321740].

\bibitem{kspysw15} D. Klochkov, V. Suleimanov, G. Puhlhofer, D. G. Yakovlev,
A. Santangelo, and K. Werner, \emph{The neutron star in HESSJ1731-347:
Central compact objectsas laboratories to study the equation of state of
superdense matter}, 2015 \emph{Astron. Astrophys.} \textbf{573} A53
[arXiv:1410.1055] [doi:10.1051/0004-6361/201424683].

\bibitem{okksy15} D. D. Ofengeim, A. D. Kaminker, D. Klochkov, V.
Suleimanov, and D. G. Yakovlev, \emph{Analysing neutron star in HESS
J1731--347 from thermal emission and cooling theory}, 2015 \emph{Mon. Not.
Roy. Astron. Soc.} \textbf{454} 2668 [arXiv:1510.00573]
[doi:10.1093/mnras/stv2204].

\bibitem{y01} D. G. Yakovlev, A.D. Kaminker, O.Y. Gnedin, P. Haensel, 
\emph{Neutrino emission from neutron stars}. 2001 \emph{Physics Reports} 
\textbf{354 }1 [astro-ph/0012122][doi:10.1016/s0370-1573(00)00131-9].

\bibitem{lp06} L.B. Leinson and A. P\'{e}rez, \emph{Vector current
conservation and neutrino emission from singlet-paired baryons in neutron
stars}, 2006 \emph{Phys. Lett.} \textbf{B 638} 114 [astro-ph/0606651]
[doi:10.1016/j.physletb.2006.05.036].

\bibitem{fm79} B. L. Friman and O. V. Maxwell, \emph{Neutrino emissivities
of neutron stars}, 1979 \emph{Astrophys. J.} \textbf{232} 541
[doi:10.1086/157313].

\bibitem{hpr01} C. Hanhart, D. R. Phillips, and S. Reddy, \emph{Neutrino and
axion emissivities of neutron stars from nucleon--nucleon scattering data},
2001 \emph{\ Phys. Lett.} \textbf{B 499} 9 [astro-ph/0003445]
[doi:10.1016/S0370-2693(00)01382-4].

\bibitem{st83} Shapiro S. L., Teukolsky S. A., 1983, \emph{Black Holes,White
Dwarfs, and Neutron Stars}. Wiley, New York

\bibitem{pa92} D. Page, J. H. Applegate, \emph{The cooling of neutron stars
by the direct URCA process}, 1992 \emph{Astrophys. J.} \textbf{394} L17
[doi:10.1086/186462].

\bibitem{yp04} D. G. Yakovlev \& C. J. Pethick, \emph{Neutron Star Cooling}
, 2004 \emph{ARA\&A} \textbf{42} 169 [astro-ph/0402143]
[doi:10.1146/annurev.astro.42.053102.134013].

\bibitem{lp01} J. M. Lattimer and M. Prakash, \emph{Neutron Star Structure
and the Equation of State}, 2001 \emph{Astrophys. J.} \textbf{550} 426
[astro-ph/0002232] [doi:10.1086/319702].

\bibitem{t39} R. C. Tolman, \emph{Static solutions of Einstein's field
equations for spheres of fluid}, 1939 \emph{Phys. Rev.} \textbf{55} 364
[doi:10.1103/PhysRev.55.364].

\bibitem{thorne77} K. S. Thorne, \emph{The relativistic equations of stellar
structure and evolution}, 1977 \emph{Astrophys. J.} \textbf{212} 825
[doi:10.1086/155108].

\bibitem{Serot} B. D. Serot and J. D. Walecka, \emph{The Relativistic
Nuclear Many-Body Problem}, 1986 \emph{Adv. Nucl. Phys.} \textbf{19}, Volume
16, eds. J.W. Negele and E. Vogt, (Plenum, New York); B. D. Serot, \emph{%
Quantum hadrodynamics}, 1992 \emph{Rep. Prog. Phys.} \textbf{55} 1855
[doi:10.1088/0034-4885/55/11/001].

\bibitem{l02} L. B. Leinson, \emph{Direct Urca processes on nucleons in
cooling neutron stars}, 2002 \emph{Nucl. Phys.} \textbf{A 707} 543
[hep-ph/0207116] [doi:10.1016/S0375-9474(02)00991-0].

\bibitem{l02plb} L. B. Leinson, \emph{Weak magnetism effects in the direct
Urca processes in cooling neutron stars}, 2002 \emph{Phys. Lett.} \textbf{B
532} 267 [hep-ph/0206097] [doi:10.1016/S0370-2693(02)01587-3].

\bibitem{gs80} G. Glen, P. Sutherland, \emph{On the cooling of neutron stars}%
, 1980 \emph{Astrophys. J.} \textbf{239} 671 [doi:10.1086/158154].

\bibitem{bpy16} M. V. Beznogov, A. Y. Potekhin, and D. G.Yakovlev, \emph{%
Diffusive heat blanketing envelopes of neutron stars},
2016 \emph{Mon. Not. R. Astron. Soc.}\textbf{459} 1569 [arXiv:1604.00538] 
[doi:10.1093/mnras/stw751].

\bibitem{plps04} Page, D., Lattimer, J. M., Prakash, M. \& Steiner, A. W., 
\emph{Minimal Cooling of Neutron Stars: A New Paradigm}, 2004 \emph{%
Astrophys. J. Supp.} \textbf{155} 623 [astro-ph/0403657]
[doi:10.1086/424844].

\bibitem{yhshp11} D.G. Yakovlev, W.C.G. Ho, P.S. Shternin, C.O. Heinke and
A.Y. Potekhin, \emph{Cooling rates of neutron stars and the young neutron
star in the Cassiopeia A supernova remnant}, 2011 \emph{Mon. Not. R. Astron.
Soc.} \textbf{411} 1977 [arXiv:1010.1154]
[doi:10.1111/j.1365-2966.2010.17827.x].

\bibitem{k92} M. Koshiba, \emph{Observational neutrino astrophysics}, 1992 
\emph{Phys. Rep.} \textbf{220} 229 [doi:10.1016/0370-1573(92)90083-C].
\end{thebibliography}
\end{document}